\documentclass[11pt,letterpaper]{article}
\usepackage[margin=1in]{geometry}
\usepackage[super,sort&compress,numbers]{natbib}

\usepackage[singlelinecheck=false]{caption}
\usepackage{setspace,array,verbatim,float,multirow}
\usepackage{ucs,amssymb,amsmath,mathtools}
\usepackage[utf8]{inputenc}
\usepackage{fontenc,graphicx}
\usepackage[usenames,pdftex,dvips]{color,xcolor}
\usepackage[pdftex,colorlinks]{hyperref,ragged2e}
\newcommand{\matr}[1]{\mathbf{#1}}
\newcommand{\Mtr}[1]{\boldsymbol{#1}}
\newcommand{\Nb}{ \nobreak}
\newcommand{\Rh}{{\mathcal R}}

\newcommand{\Pv}{P\nobreakdash-value}

\newcommand{\Pvs}{P\nobreakdash-values}
\title{DOT: Gene-set analysis by combining decorrelated association statistics
}
\author{Olga A. Vsevolozhskaya,$^1$ Min Shi,$^2$ Fengjiao Hu,$^2$ Dmitri V. Zaykin$^{2\,\ast}$}

\begin{document}
\doublespacing
\date{}
\maketitle
\thispagestyle{empty}

\noindent
$^1$Department of Biostatistics, College of Public Health, University of Kentucky, Lexington, KY 40536, USA. \\
\noindent
$^2$Biostatistics and Computational Biology, National Institute of Environmental Health Sciences, National Institutes of Health, Research Triangle Park, NC 27709, USA.\\
\vskip 2ex
\noindent
$^\ast$Correspondence: Dmitri V. Zaykin, Senior Investigator at the Biostatistics and Computational Biology Branch, National Institute of Environmental Health Sciences, National Institutes of Health, P.O. Box 12233, Research Triangle Park, NC 27709, USA. Tel.: +1 (919) 541-0096 ; Fax: +1  (919) 541-4311. Email address: dmitri.zaykin@nih.gov
\clearpage

\section*{Abstract}
Historically, the majority of statistical association methods have been designed assuming availability of SNP-level information. However, modern genetic and sequencing data present new challenges to access and sharing of genotype-phenotype datasets, including cost management, difficulties in consolidation of records across research groups, etc. These issues make methods based on SNP-level summary statistics particularly appealing. The most common form of combining statistics is a sum of SNP-level squared scores, possibly weighted, as in burden tests for rare variants. The overall significance of the resulting statistic is evaluated using its distribution under the null hypothesis. Here, we demonstrate that this basic approach can be substantially improved by decorrelating scores prior to their addition, resulting in remarkable power gains in situations that are most commonly encountered in practice; namely, under heterogeneity of effect sizes and diversity between pairwise LD. In these situations, the power of the traditional test, based on the added squared scores, quickly reaches a ceiling, as the number of variants increases. Thus, the traditional approach does not benefit from information potentially contained in any additional SNPs, while our decorrelation by orthogonal transformation (DOT) method yields steady gain in power. We present theoretical and computational analyses of both approaches, and reveal causes behind sometimes dramatic difference in their respective powers. We showcase DOT by analyzing breast cancer data, in which our method strengthened levels of previously reported associations and implied the possibility of multiple new alleles that jointly confer breast cancer risk.

\clearpage
\section*{Introduction}
During the recent years, genome-wide association studies (GWAS) uncovered a wealth of genetic susceptibility variants. The emergence of new statistical approaches for the analysis of GWAS have largely contributed to that success. The majority of these methods require access to individual-level data, yet methods that require only summary statistics have been developed as well. The rising popularity of summary-based methods for the analysis of genetic associations has been motivated by many factors, among which is convenience and availability of summary statistics and high statistical power that can often match the power of analysis based on individual records.\cite{DanyuLinMetaNoGain2009,lee2013general,zaykin2011optimally}

Many types of association tests, including those originally developed for individual-level records, can be presented in terms of added summary statistics. For example, gene set analysis (GSA) tests or burden and overdispersion tests for rare variants,\cite{pasaniuc2017dissecting,lee2013general,Gates_ajhg} can be written as a weighted sum of summary statistics. In GSA applications, methods based on combined summary statistics can be used to efficiently aggregate information across many potentially associated variants within individual genes, as well as over several genes that may represent a common etiological pathway. When within-gene association statistics (or equivalently, \Pvs{}) are being combined, linkage disequilibrium (LD) needs to be accounted for, because LD induces correlation among statistics. The correlation among association test statistics for individual SNPs without covariates is the same as the correlation between alleles at the corresponding SNPs. This fact allows one to model a set of statistics using a multivariate normal (MVN) distribution with the correlation matrix equal to the matrix of LD correlations. More generally, in the presence of covariates correlated with SNPs, MVN correlations among association statistics will depend not only on LD but also on other covariates in the model.\cite{conneely20071158,Sun361436}

When SNPs are coded as 0,1,2 values, reflecting the number of copies of the minor allele, the LD matrix of correlations can be obtained from SNP data as the sample correlation matrix. It can also be directly estimated from haplotype frequencies whenever those are available or reported. Specifically, the LD (i.e., the covariance between alleles $i$ and $j$; $D_{ij}$) is defined by the difference between the di-locus haplotype frequency, $P_{ij}$, and the product of the frequencies of two alleles, $D_{ij} = P_{ij} - p_{i}p_{j}$. Then, the correlation between a pair of SNPs is defined as $r_{ij} = \frac{D_{ij}}{p_i(1-p_i)p_j(1-p_j)}$.   The di-locus $P_{ij}$ frequency is defined as the sum of frequencies of those haplotypes that carry both of the minor alleles for SNPs $i$ and $j$. Similarly, $p_i$ allele frequency is the sum of haplotype frequencies that carry the minor allele of SNP $i$.

It is important to distinguish situations, in which the LD matrix is estimated using the same data that was used to compute the association statistics from those, in which the estimated LD matrix is obtained based on a suitable population reference panel.  The reference panel approach is implemented in popular web-based association analysis platforms, such as ``VEGAS''\cite{liu2010} or ``Pascal.''\cite{lamparter2016fast} Based on a user-provided list of $L$ SNPs, with the corresponding association \Pvs{}, VEGAS queries an online reference panel resource to obtain the matrix of LD correlations. \Pvs{} are then transformed to normal scores $P_i \rightarrow Z_i$, $i=1,\dots,L$, and vector $\matr{Z}$ is assumed to follow zero-mean MVN distribution under the null hypothesis of no association. The individual statistics in VEGAS are then combined as  $\text{TQ} = \sum_{i=1}^L Z_i^2$, (where $\text{TQ}$ stands for ``Test by Quadratic form'') and the overall SNP-set \Pv{} is derived empirically by simulating a large number ($j=1,\dots,B$) of zero-mean MVN vectors, adding their squared values to obtain statistics $\text{TQ}^{(j)}$ and computing the proportion of times when $\text{TQ}^{(j)} > \text{TQ}$. The statistics similar to $\text{TQ}$ are ubiquitous and appear in many 
proposed tests that aggregate association signals within a genetic region.

As exemplified by VEGAS, the distribution of $\text{TQ}$ must explicitly incorporate LD. However, an alternative approach that implicitly incorporates LD can be based on first decorrelating the association summary statistics,  and then exploiting the resulting independence to evaluate the distribution of the sum of decorrelated statistics, which we call Decorrelation by Orthogonal Transformation ($\text{DOT}$). This general idea is straightforward and have been used in many contexts. For instance, Zaykin et al. suggested a variation of this approach for combining \Pvs{} (or summary statistics) but have not studied power properties of the method in detail.\cite{Zaykin2002}

Here, we propose a new decorrelation-based method. We derive theoretical properties of our method and explore asymptotic power of both $\text{DOT}$ and $\text{TQ}$ type of statistics. To the best of our knowledge, we are the first ones to derive the asymptotic distributions of $\text{DOT}$ and $\text{TQ}$ under the alternative hypothesis. Our results show that decorrelation can provide surprisingly large power boost in biologically realistic scenarios. However, high statistical power is not the only advantage of the proposed framework. Once statistics are decorrelated, one can tap into a wealth of powerful methods developed for combining independent statistics. 
These methods, among others, include 
approaches that emphasize the strongest signals by combining the top-ranked results.\cite{dudbridge2003,Zaykin2002,zaykin2007combining,biernacka2012use,fridley2013soft,taylor2005tail}

Our theoretical analyses also reveal an unexpected result, showing that in many practical settings tests based on the statistic $\text{TQ}$ do not gain power with the increase in $L$ (assuming the same pattern of effect sizes for different values of $L$), while the proposed method steadily gains power  under the same conditions. Specifically, the proposed decorrelation method gains power when the effect sizes and/or pairwise LD values become increasingly more heterogeneous.  The reasons behind the respective behaviors of tests based on $\text{TQ}$ and $\text{DOT}$ are explored here theoretically and confirmed via simulations. We further derive power approximations that are useful for understanding power properties of the studied methods.

To showcase our method, we evaluate associations between breast cancer susceptibility and SNPs in estrogen receptor alpha (\textit{ESR1}), fibroblast growth factor receptor 2 (\textit{FGFR2}), RAD51 homolog B (\textit{RAD51B}), and TOX high mobility group box family member 3 (\textit{TOX3}) genes, without access to raw genotype data. We first test for a joint association between SNPs in those four genes and breast cancer risk by decorrelating summary statistics based on the overall LD gene structure. We then describe how to follow up on the joint association results and identify one or more SNPs that drive joint association with disease risk. Our study confirms previous associations and reveals new associations, suggesting new potential breast cancer SNP markers.

\section*{Material and Methods}
Genetic association tests based on summary statistics are often presented as a weighted sum.\cite{pasaniuc2017dissecting,lee2013general} Let $w_i$ denote the weight assigned to individual statistic. The weighted statistics can then be defined as $Y^2_i\Nb=\Nb w_iZ^2_i$ with $\matr{Z}\Nb\sim\Nb\text{MVN}(\boldsymbol{\mu}_Z,\matr{\Sigma}_Z)$ and
$\matr{Y}\Nb\sim\Nb\text{MVN}(\boldsymbol{\mu},\matr{\Sigma})$, where $\Mtr{\mu}=\matr{W}\boldsymbol{\mu}_Z$, $\matr{\Sigma} = \matr{W} \matr{\Sigma}_Z \matr{W}$, and $\matr{W} = \text{diag}(\matr{\sqrt{w}})$. The statistics $Y_i^2$ are marginally distributed as one degree of freedom chi-square variables with noncentralities $\mu_i^2$. The overall statistic is then typically defined as $\text{TQ} = \sum_{i=1}^L Y_i^2$. 
\subsection*{Joint distribution of association summary statistics}
In this section, we derive parameters $\boldsymbol{\mu}$ and $\matr{\Sigma}$ of the joint MVN distribution of summary statistics. Under the null hypothesis, when none of the SNPs are associated with an outcome, $\boldsymbol{\mu} = \boldsymbol{0}$. If individual SNP models do not include covariates, $\matr{\Sigma}_Z$ equals the LD matrix, i.e., the correlation matrix between the SNP values coded as 0, 1, or 2,  reflecting the number of minor alleles in a genotype. In the presence of covariates, $\matr{\Sigma}_Z$ is a Schur complement of the submatrix of the matrix of all predictor variables.\cite{conneely20071158} That is, the estimated correlation between association statistics $\hat{\matr{\Sigma}}_Z$ can be obtained by inverting the covariance or correlation matrix of all predictors, selecting the SNP submatrix, inverting it back, and standardizing  the result to correlation.

Under the alternative hypothesis, when some SNPs are associated with a trait $y$, let $\beta_j$ be the regression coefficient for the $j$-th SNP. Then, a typical linear model that determines the trait value is defined as:
\begin{eqnarray}
   y = \beta_0 + \sum_{j=1}^L \beta_j \, \text{SNP}_j + \epsilon, \label{linear.model}
\end{eqnarray}
where $\epsilon \sim N(0,1)$. The mean value of the summary statistics (i.e., noncentralities) can be expressed as:
\begin{eqnarray}
   \mu_j = \sqrt{N} \frac{\matr{\Sigma}_{j}\Mtr{\beta} }{ \Mtr{\beta}'\matr{\Sigma}\Mtr{\beta} + 1} = \sqrt{N} \,\, b_j, \label{nonc.linear}
\end{eqnarray}
where $\matr{\Sigma}_{j}$ is the $j$-th column of $\matr{\Sigma}$, $b_j=  \text{cor}(y, \text{SNP}_j)$ and $N$ is the sample size. We note that Eq. (\ref{nonc.linear}) is valid outside of the linear model settings. For example, consider a latent variable model, where the continuous unobserved (latent) variable $y_l$ is linear in predictors according to Eq. (\ref{linear.model}), and the observed variable (disease status) is $y=1$ whenever $y_l > l$ and $y=0$ otherwise. When such binary outcome is analyzed by logistic regression, a good approximation to the noncentrality values will be:
\begin{eqnarray}
   \mu_j \approx \sqrt{N} (d  \times b_j). \label{nonc.nonlinear}
\end{eqnarray}
If error terms $\epsilon$ are assumed to be normally distributed, the reduction in correlation due to dichotomization by the factor $d$ can be expressed as $d = \phi(l) / \sqrt{\Phi(l)(1-\Phi(l))}$, where $\phi(\cdot), \Phi(\cdot)$ are the probability and the cumulative densities of the standard normal distribution.\cite{maccallum2002practice}

Under association, surprisingly, the correlation matrix between statistics is no longer $\matr{\Sigma}$. Let $\sigma_{ij}$ be the $i,j$-th element of  $\matr{\Sigma}$. By using the multivariate delta method, we derived the $i,j$-th element of the correlation matrix $\Mtr{\Rh}$ as follows:
\begin{eqnarray}
\rho_{ij} &\approx& \frac{(\mu_i \mu_j (\mu_i^2 + \mu_j^2 - N) - 2 (\mu_i^2 + \mu_j^2 - N) N \sigma_{ij} + 
  \mu_i \mu_j N \sigma_{ij}^2) } {(\mu_i^2 - N) (\mu_j^2 - N)}, \label{induced.cor.1} \\
          &\approx&   \rho_{ij} +  \rho_{ij}\frac{\mu_i\mu_j}{2 N} - \rho_{ij}\frac{\mu_i^2\mu_j^2}{N^2} - \frac{\mu_i\mu_j}{2 N}, \nonumber \\
  &=&   \rho_{ij} +  \rho_{ij} b_i b_j - \rho_{ij}b_i^2 b_j^2 - b_i b_j.
                    \label{induced.cor}
\end{eqnarray}
Note that when some of SNP pairs ($i,j$) are associated, summary statistics may become correlated even if there is no LD between the SNPs, due to the last term, $-b_i b_j$, in Eq. (\ref{induced.cor}). Equations (\ref{nonc.linear}), (\ref{nonc.nonlinear}), (\ref{induced.cor.1}), (\ref{induced.cor}) allow one to study power properties of the methods based on sums of association statistics, as well as to design realistic simulation experiments, where summary statistics can be sampled directly from the MVN distribution under the alternative hypothesis. That is, given effect sizes and the correlation matrix among predictors, statistics can be immediately sampled from the MVN$(\Mtr{\mu}, \Mtr{\Rh})$ distribution. This approach avoids both the data-generating step and the subsequent computation of summary statistics from that data, leading to a substantial gain in computation time. In certain situations, the difference in speed can be dramatic. For example, it is not trivial to simulate discrete (genotype) data given a specific LD matrix. Current state of the art methods tend to be slow, because they rely on ad hoc iterative techniques, such as generation of multiple random ``proposal'' data sets to fit the target correlation matrix.\cite{ferrari2012simulating}


Results of simulation experiments presented here were performed based on effect sizes specified via the linear model (Eq. \ref{linear.model}). However, we verified (not presented here) the validity of the proposed theory assuming logistic, probit, and Poisson regression models. We also note that Conneely et al. presented theoretical arguments supporting the validity of the MVN joint distribution of summary statistics under no association for a broad class of generalized regression models.\cite{conneely20071158}

\subsection*{Distribution of sums of association summary statistics}
As we noted at the beginning of the ``Methods'' section, weighted sums of summary statistics can be re-expressed as unweighted sums, where the mean and the correlation parameters are modified to absorb the weights. The distribution of  $\sum_{i=1}^LY_i^2$ follows the weighted sum of independent one degree of freedom non-central chi-square random variables. Although this result is standard, the components of this weighted sum depend on the joint distribution of association summary statistics under the alternative hypothesis, and this distribution has not been previously derived. In the previous section, we provide the components of $\boldsymbol{\mu}$ and $\Mtr{\Rh}$ that determine the weights and the noncentralities of chi-squares. Therefore,
\begin{eqnarray}
  \Pr\left(\matr{Y}'\matr{Y} > t \right) &=& \Pr\left(\sum_{i=1}^L Y_i^2 > t \right) = \Pr\left(\sum_{i=1}^L \lambda_i \chi^2_{1, \gamma_i} > t \right), \label{yy} \\
  \boldsymbol{\gamma} &=& \left\{\boldsymbol{\mu}' \, \matr{E} \, \left(\frac{1}{\sqrt{\boldsymbol{\lambda}}} \, \matr{I} \right)\right\}^2,
\end{eqnarray}
where the weights, $\boldsymbol{\lambda}$, are the eigenvalues of $\Mtr{\Rh}$ and $\boldsymbol{\gamma}$ is the vector of non-centrality parameters. The columns of the matrix $\matr{E}$ are orthogonalized and normalized eigenvectors of $\Mtr{\Rh}$. The \Pv{} for the statistic $\text{TQ}=\matr{Y}'\matr{Y}$ is obtained by setting $\boldsymbol{\mu}$ to zero and then calculating this tail probability at the observed value $\text{TQ}=t$. Note that the elements in $\Mtr{\Rh}$, and therefore the eigenvalues $\lambda_i$, explicitly depend on the $\beta$-coefficients through Eqs. (\ref{nonc.linear}) and (\ref{induced.cor}).

Our decorrelation approach uses a symmetric orthogonal transformation of the vector of statistics $\matr{Y}$ to a new vector $\matr{X}$, with the new joint statistic based on the sum of elements of $\matr{X}$, $\text{DOT}=\sum_{i=1}^L X^2_i$. The orthogonal transformation is defined as follows. Let $\matr{D} = \left(\frac{1}{\sqrt{\boldsymbol{\lambda}}} \, \matr{I} \right)$ and define $\matr{X} =  \matr{H}\; \matr{Y}$, where $\matr{H} = \matr{E} \, \matr{D} \matr{E}'$. The squared values, $X_i^2$, are one degree of freedom independent chi-square variables, thus $\text{DOT} = \matr{X}'\matr{X}$ is a chi-square random variable with $L$ degrees of freedom and noncentrality value of:
\begin{eqnarray}
   \gamma_c &=& \sum_{i=1}^L \gamma_i = \boldsymbol{\mu}'\Mtr{\Rh}^{-1} \boldsymbol{\mu} = \left( \matr{H}\; \boldsymbol{\mu}\right)' \left( \matr{H} \, \boldsymbol{\mu} \right) \label{gamma.c}.
\end{eqnarray}
The cumulative distribution of the new test statistic is thus,
\begin{eqnarray}
   \Pr\left(\matr{X}'\matr{X} > t\right) = \Pr\left(\chi^2_{L,\gamma_c} > t\right). \label{xx}
\end{eqnarray}

There are many ways to choose an orthogonal transformation, but a valid one for our purposes needs to have the following ``invariance to order'' property. Suppose we sample an equicorrelated MVN vector $\matr{Y}$ with a common correlation $\rho$ for all pairs of variables. Before decorrelating the vector, we permute its values to a different order. A permutation in this example is a legitimate operation, because an equicorrelation structure does not suggest a particular order of $\matr{Y}$ values. After an orthogonal transformation of  $\matr{Y}$ to  $\matr{X}$, the order of $\matr{X}$ entries may change due to permutation but their values should remain the same. Moreover, for the method to be useful in practice, we need the invariance to hold for a more general class of statistics than a simple sum of chi-squares, $\sum^L_{i=1} X^2_i$. For example, the Rank Truncated Product (RTP) is a powerful \Pv{} combination  method\cite{dudbridge2003} that emphasizes small \Pvs{}: the RTP statistic $T_{\text{RTP}}$ is the product of the $k$ smallest \Pvs{}, $k < L$, or equivalently, $T_{\text{RTP}}=\sum^k_{i=1}\left[-\ln(P_i)\right]$, where $P_1 \le P_2 \dots \le P_k$. Note that $-\ln(P_i)$ is no longer a one degree of freedom chi-square variable.

The ``invariance to order'' requirement implies that the value of DOT-statistic should not change due to a permutation of (equicorrelated) values in $\matr{Y}$. Not all orthogonal transformations meet the invariance to order criteria. It can be easily verified that neither the inverse Cholesky factor ($\matr{C}^{-1}$) transformation, $\matr{X}=\matr{C}^{-1}\;\matr{Y}$, nor another commonly used transformation $\matr{X} = \matr{E} \, \left(\frac{1}{\sqrt{\boldsymbol{\lambda}}} \, \matr{I}\, \right)\matr{Y}$, have the invariance to order property, except in the special case of the sum of $L$ chi-squared variables $\sum_{i=1}^L X_i^2$. 
To clarify, we call this statistic ``the special case,'' because, for example, in the case of RTP with $k=L$, the statistic $\sum_{i=1}^L -\ln(P_i)$ is no longer the sum of one degree of freedom chi-squares. Moreover, some transformations of equicorrelated data to independence, such as the Helmert transformation, may change values of $\matr{X}$ depending on the order of values in $\matr{Y}$, even in a special equicorrelation case of $\rho=0$ (i.e., when variables in $\matr{Y}$ are independent). The proposed $\mathbf{H}$, as defined above, has both the invariance to order property and can be used with \Pv{} transformations other than that to the one degree of freedom chi-square.

\subsection*{Theoretical analysis of power}

For exploration of power properties, it is useful to first consider the equicorrelation case, because in this case it is possible to derive illustrative equations that relate power to: (1) the number of SNPs, $L$; (2) the common correlation value for every pair of SNPs, $\rho$; and (3) the mean values of association statistics, $\boldsymbol{\mu}$. In the equicorrelation case, the correlation matrix can be expressed as $\Mtr{\Rh}_\rho = (1-\rho) \matr{I} + \rho \matr{1 1'}$. The eigenvalue vector of $\Mtr{\Rh}_\rho$ has length $L$ but only two distinct values, $\boldsymbol{\lambda} = \{1+\rho(L-1),1-\rho,\dots,1-\rho\}$.

For decorrelated statistic $\text{DOT}$, we derived a simple form of $L$ noncentralities by utilizing the Helmert orthogonal eigenvectors\cite{clarkehelmert,lancaster1965} as follows:
\begin{eqnarray}
  \delta_1 &=& \frac{(\sum_{i=1}^L{\mu_i})^2}{L\left(1+(L-1)\rho\right)} = \frac{L \, \bar{\boldsymbol{\mu}}^2}{1+(L-1)\rho}, \label{delta.1.equi} \\
  \delta_{j>1} &=& \sum_{i=1}^{j-1} \frac{\left(\mu_i-\mu_j\right)^2}{L (1-\rho)}, \label{delta.j.equi} 
\end{eqnarray}
where $\bar{\boldsymbol{\mu}}$ is the average of the values in $\boldsymbol{\mu}$. 
Next, let
\begin{eqnarray}
   \delta_s = \sum_{j=2}^L \delta_j = (L-1) \frac{\bar{d}}{2(1-\rho)}, \label{deltasum}
\end{eqnarray}
where $\bar{d}$ is the average of $d_{ij}=\left(\mu_i-\mu_j\right)^2$, over all pairs of $\mu_i$ and $\mu_j$, such that $i<j$. The values in $d_{ij}$ are the pairwise squared differences in the standardized effect values as captured by the vector $\boldsymbol{\mu}$. This representation yields the noncentrality of $\text{DOT}$ as a function of the common correlation and the mean standardized effect size as:
\begin{eqnarray}
   \gamma_c =  \frac{L \, \bar{\boldsymbol{\mu}}^2}{1+(L-1)\rho} +  \delta_s  \label{gamma.c.equi}.
\end{eqnarray}
Note that as $L$ increases, the first term in Eq. (\ref{gamma.c.equi}) approaches $\bar{\boldsymbol{\mu}}^2/\rho$, while the sum of the remaining noncentralities, $\delta_s$, increases linearly with $L$, as long as the average of the squared effect size differences, $\overline{d}$, does not depend on $L$. Thus, the noncentrality of the decorrelated statistic $\text{DOT}$ is expected to steadily increase with $L$ and become approximately $\bar{\boldsymbol{\mu}}^2/\rho + (L-1)\frac{\bar{d}}{2(1-\rho)}$.

Next, we consider the distribution of the statistic $\text{TQ}=\matr{Y}'\matr{Y}$.  Note that $\sum_{i=1}^L \delta_i = \sum_{i=1}^L \gamma_i$, where $\gamma_i$'s are the noncentralities for $\text{TQ}$ and $\delta_i$'s are the noncentralities of $\text{DOT}$. In the equicorrelation case, the distribution $\text{TQ}$ reduces to the weighted sum of two chi-square variables, because there are only two distinct eigenvalues that correspond to $\Mtr{\Rh}_\rho$, namely:
\begin{eqnarray}
  \Pr\left(\matr{Y}'\matr{Y} > t \right) &=& \Pr\left\{ (1+(L-1)\rho) \chi^2_{1, \gamma_1}
  + (1-\rho) \chi^2_{L-1, \gamma_c - \gamma_1} > t \right\} \\
  &=& \Pr\left\{  \chi^2_{1, \gamma_1} + \frac{1-\rho}{1+(L-1)\rho} \chi^2_{L-1, \gamma_c - \gamma_1} > \frac{t}{1+(L-1)\rho} \right\}.\label{sec}
\end{eqnarray}
The term $\frac{1-\rho}{1+(L-1)\rho} \chi^2_{L-1, \gamma_c - \gamma_1}$ in Eq. (\ref{sec}) approaches the constant $\frac{\bar{d}(1-\rho)}{2 \rho^2}\Nb+\Nb\frac{1-\rho}{\rho}$ as $L$ increases. Therefore, under the null hypothesis, the distribution of the quadratic form $\matr{Y}'\matr{Y}$ can be well approximated by the location-scale transformation of the one degree of freedom chi-squared random variable:
\begin{eqnarray}
   \Pr \left\{ \frac{\matr{Y}'\matr{Y} - (L-1)(1-\rho)}{(L-1)\rho +1} > \chi^2_{\alpha} \right\} \approx \alpha \label{approxyy},
\end{eqnarray}
where $\chi^2_\alpha$ is $1-\alpha$ quantile of the one degree of freedom chi-square distribution.

To summarize, we just showed that the distribution of the decorrelated set of variables gains in the total noncentrality with $L$, while the distribution of the sum $\matr{Y}'\matr{Y}$ depends heavily only on the noncentrality of the first term, $\gamma_1$. The approximate power of the test based on the statistic $\text{TQ}=\matr{Y}'\matr{Y}$ can be computed as:
\begin{eqnarray}
   \Pr(\text{TQ} > t) &\approx& 1-\Psi\left(t\right), \label{approxyy1} \\
   t &=& \chi^2_\alpha + \frac{1-\rho^\ast}{\rho^\ast} + \frac{1}{2} \frac{(1-\rho^\ast) \bar{d}}{(\rho^\ast)^2},
\end{eqnarray}
where $\rho^\ast=\sqrt{\overline{\rho_{ij}^2}}$, $\mu^\ast = \overline{|\boldsymbol{\mu}|}$
and $\Psi(\cdot)$ is a one degree of freedom chi-square CDF with the noncentrality $L \mu^\ast / ((L-1) \rho^\ast + 1)$, evaluated at $t$. The ceiling noncentrality value, as $L\rightarrow \infty$, is thus $\mu^\ast/\rho^\ast$. Let us re-emphasize the point that a test based on the distribution of the $\text{TQ}$ statistic is expected to be less powerful than $\text{DOT}$ in the presence of heterogeneity among effect sizes. Heterogeneity in LD will contribute to the difference in power. Starting with an equicorrelation model, we can introduce perturbations to the common value, $\rho>0$, by adding noise derived from a rank-one matrix $\matr{U}\matr{U}'$, where $\matr{U}$ is a vector of random numbers. Specifically, perturbations can be added as $\matr{B} = \Mtr{\Rh}_\rho + \matr{U}\matr{U}'$. Next, $\matr{B}$ should be standardized to correlation as $\matr{B}_R\Nb=\Nb\left\{\sqrt{1/\text{Diag}(\matr{B})} \, \matr{I} \right\}\, \matr{B}
\left\{\sqrt{1/\text{Diag}(\matr{B})} \, \matr{I} \right\}$. When elements in $\matr{U}$ are close to zero, the matrix $\matr{B}_R$ deviates from $\Mtr{\Rh}_\rho$ by only a small jiggle around $\rho$. 
Matrix $\matr{B}_R$ provides a way to construct random correlation matrices in a controlled manner, where the degree of departure from the equicorrelation is controlled via the range of the elements in $\matr{U}$. The utility of $\matr{B}_R$ is that it represents a perturbation of $\Mtr{\Rh}_\rho$, and we expect our power results under equicorrelation case to hold approximately, at least for small jiggles around $\rho$. Nevertheless, it turns out that even for a more general correlation structure, our power approximations still hold, which we show via extensive simulation studies.

\subsection*{LD patterns from the 1000 Genome Project}
In a separate set of simulation experiments, we utilized realistic LD patterns using data from the 1000 Genomes Project.\cite{10002012integrated} For every simulation experiment, we selected a random set of consecutive SNPs from a chromosome 17 region, that was spanning over 100 Kb and included SNPs from the gene \textit{FGF11} to the gene \textit{NDEL1}. Each stretch of consecutive SNPs contained from 10 to 200 SNPs with the minimum allele frequency 0.025. A random portion of SNPs in every set carried no effect on the outcome on its own, and we considered these SNPs to be ``proxies'' for causal variants due to LD. The median LD correlation varied from approximately -0.6 to 0.98 between random stretches of SNPs. The number of proxy SNPs varied from 3 to 197 across simulations. The sample size was also set to be random and varied from 500 to 3000 across simulations. Effect sizes for causal variants were modeled by $\beta$-coefficients, as given by Eq. (\ref{linear.model}), and drawn randomly from the interval [-0.4, 0.4]. Summary statistics were sampled from the MVN distribution with parameters given by Eqs. (\ref{nonc.linear}), (\ref{induced.cor.1}).
To check the validity of our approach of sampling the summary statistics directly, we first conducted a separate set of extensive simulation experiments, in which power and type-I error rates were obtained by simulating individual data and then $\text{TQ}$ and $\text{DOT}$ statistics were computed by running the actual regression analysis. We confirmed excellent agreement between the two approaches, thus most of the subsequent simulations were conducted by sampling the summary statistics directly (these results are not shown here).

\section*{Results}
We conducted simulation experiments to study statistical power of the proposed method based on the decorrelation statistic $\text{DOT}$, and to compare it to the statistic $\text{TQ}$. We also included a recently proposed method ``ACAT'' by Liu and colleagues,\cite{liu2019acat} where association \Pvs{} for individual SNPs are transformed to Cauchy-distributed random variables, then added up to obtain the overall \Pv{}. ACAT was included into comparisons because it has robust power across different models of association. Specifically, Liu et al. found ACAT to be competitive against popular methods, including SKAT and burden tests for rare-variant associations.\cite{wu2011rare,li2008methods,madsen2009groupwise,price2010pooled} A distinctive feature of ACAT is its good type-I error control in the presence of correlation between \Pvs{}, which, interestingly,  improves as the $\alpha$-level becomes smaller, due to its usage of transformation to a moment-free Cauchy distribution.

We used two distinct scenarios in our simulation experiments:
\begin{enumerate}
\item First, we assumed that the summary statistics and the sample correlation matrix among statistics are estimated from the same data set. This allowed us to validate power properties derived in ``Methods.'' 
\item Second, we assumed that the sample correlation LD matrix was obtained from external reference panel. We included this scenario into our simulations due to the concern that the type-I error rate of the methods considered here may be inflated if the correlation matrix is computed based on a separate data set.
\end{enumerate}
\subsection*{Simulations assuming that the LD matrix and the summary statistics are obtained from the same data}
To compare methods with and without decorrelation of statistics, we considered several distinct settings. 
\begin{itemize}
\item[Setting 1.]  The decorrelation method ($\text{DOT}$) is expected to gain power as the number of SNPs increases in scenarios where effect sizes vary markedly from SNP to SNP. However, if effect sizes for all SNPs are in fact very close to each other, the power of $\text{DOT}$ may decrease. To illustrate this property, our first, and purposely contrived simulation setup is where the effect sizes were all non-zero but very close to each other in their magnitude, varying uniformly from 2.3 to 2.4 (these are the values of the means of normally distributed standardized statistics). Table \ref{tab1} shows the results of the simulations study under this setting, in which the decorrelation method was deliberately set up to fail. In the table, the columns labeled ``Theoretic.'' provide power calculated based on the distribution of the test statistics under the alternative hypothesis that we derived above. The columns labeled  ``Empiric.'' provide results based on the empirical evaluation of power by computing P-values under the null. The columns labeled ``Approx.'' provide power calculated based on the Eq. (\ref{approxyy1}). The column labeled $\bar{\gamma}$ provide the average noncentrality value.

  The table illustrates that our analytical calculations under the alternative hypothesis are correct. That is, the empirical power of both $\text{TQ}$ and $\text{DOT}$ statistics matches nearly exactly the analytical calculations. The approximation based on Eq. (\ref{approxyy1}) apparently works well as well, emphasizing the fact that the distribution of the $\text{TQ}$ statistic can be well approximated by a one-degree of freedom chi-square distribution.

  Further, the table confirms that the decorrelation method is under-performing relative to $\text{TQ}$ if there is very little heterogeneity among SNP effect sizes. Nonetheless, this scenario is admittedly unrealistic in practice. Furthermore, the table also illustrates that as the average non-centrality value increases, the power of $\text{DOT}$ increases as well, while the power of $\text{TQ}$ is relatively constant and about 80\%. Finally, Table \ref{tab1} shows that the power of $\text{TQ}$ (although higher than that of $\text{DOT}$) does not change with $L$, highlighting the ceiling property of this method and the fact that combining more SNPs would not lead to higher power of $\text{TQ}$.

  
\item[Setting 2.] One of the features of the decorrelation method is that it benefits from heterogeneity in pairwise LD. To illustrate this property, we added jiggle to the equicorrelation matrix as described in the ``Methods'' section, while keeping the effect size vector the same as in Setting 1 (within the range of 2.3 to 2.4). In this second set of simulations, uniformly distributed perturbations (in the range 0 to 5) were added through $\matr{U}$, which made the pairwise correlations range from 0.14 to 0.98.

  Table \ref{tab2} summarizes the results and once again, illustrates the ceiling feature of $\text{TQ}$ power. However, the power of the statistic $\text{DOT}$ now starts to climb up with $L$ and the proposed test based on $\text{DOT}$ eventually becomes more powerful than the one based on $\text{TQ}$. Moreover, note that although the average noncentrality value does not increase with $L$, the $\text{DOT}$-test still gains power with $L\,\,$!


\item[Setting 3.] This setting is analogous to the equicorrelation scenario in Setting 1, except that the mean values were lowered: in Setting 1 the range in $\boldsymbol{\mu}$ was 2.3 to 2.4, while here, the range was set to vary uniformly between 1 and 2.3. Thus, the maximum effect size was lower than that in the previous simulations but the heterogeneity among effect sizes was higher. We emphasize again that while the equicorrelation assumption is unrealistic, it serves as a very useful benchmark scenario that highlights power behavior and features of the statistics $\text{TQ}$ and $\text{DOT}$ and  allows one to introduce departures from equicorrelation in a controlled manner.

  Table \ref{tab3} presents the results. The ``Approx.'' column in this table was removed and replaced by power values based on a ``\Pv{}''-approximation to the distribution of $\text{TQ}$ as in Eq. (\ref{approxyy}). This switch highlights the idea that both the power and the P-value for the $\text{TQ}$ test can be reliably estimated based on the one degree of freedom chi-squared approximation. Importantly, Table \ref{tab3} demonstrates that the power of the $\text{DOT}$-test reaches 100\% as $L$ increases (despite the fact that effect sizes were lower than in the previous settings), while the power of the $\text{TQ}$-test stays in the range 51.2 to 52.5\%, 




\item[Setting 4.] This setting is similar to the scenario in Setting 2, except that we allowed higher heterogeneity in pair-wise LD values. LD was constructed as perturbation of $\Mtr{\Rh}_{\rho=0.7} + \matr{U}\matr{U}'$ (as described in Methods), with $\matr{U}$ set to be a random sequence on the interval from -5 to 5. This resulted in LD values ranging from -0.93 to 0.99. The effect sizes were sampled randomly within each simulation from (-0.15, 0.15) interval.

  Table \ref{tab4} presents the results and shows that in this setting, the power of $\text{DOT}$ is dramatically higher than that of $\text{TQ}$ and ACAT. In fact, power values for the $\text{TQ}$ and ACAT tests barely exceed the type-I error, while the power of the decorrelation method steadily increases with $L$, eventually exceeding 90\%.

\item[Settings 5--7.] In these sets of simulations we used biologically realistic patterns of LD. Also, rather than specifying mean values of association statistics directly, we utilized a regression model for the effect sizes, as described in Eqs. (\ref{linear.model}) and (\ref{nonc.linear}). We re-iterate that when association of SNPs with a trait is present (under the alternative hypothesis), the correlation among statistics is not equal to LD, because it also has to incorporate effect sizes, as illustrated by Eq. (\ref{induced.cor}). This point is important if one wants to simulate statistics directly from the MVN distribution rather than computing them based on simulated data followed by regression.

  The results are presented in Table \ref{tab5}. Columns labeled ``Regr.'' represent scenarios, in which data were generated and statistics were computed. Columns labeled ``MVN'' represent scenarios, in which statistics were simulated directly. The rows of Table \ref{tab5} show  power values for three different $\alpha$-levels. We expected the power values in ``Regr.'' and ``MVN'' columns to match, and they do, highlighting another utility of our analytical derivation of the distribution of the test statistic under the alternative hypothesis. That is, using our results, one can significantly reduce computational and programming burden in genetic simulations. Also note that power values in Table \ref{tab5} do not decrease as $\alpha$-level becomes smaller (Settings 6 and 7). This is due to the fact that we deliberately discarded effect size and LD configurations where power was expected to be too low, because we wanted to assure a good range of power values across methods.


  As in previous simulations, power values of $\text{TQ}$ and ACAT are similar. The power approximation by Eq. (\ref{approxyy1}) remains close to the predicted theoretical power, as well as to empirically estimated powers. We also observed that power of the decorrelation test, $\text{DOT}$, is substantially higher than the powers of either $\text{TQ}$ or ACAT.
\end{itemize}
Patterns of LD and effect sizes in Settings 1--4 are not necessarily realistic biologically, however, they serve as benchmark scenarios that help to understand and highlight differences in the respective statistical power of the methods. Simulations for Settings 1--4 were performed at the 5\% $\alpha$-level based on $2\times10^6$ evaluations. Settings 5--7 used realistic patters of LD derived from the 1000 Genomes Project data. Test sizes varied from 0.001 to $10^{-7}$ with at least 10,000 simulations for power estimates. Type-I error rates were well controlled for $\text{TQ}$ and $\text{DOT}$. However, as noted by Liu et al., because the ACAT \Pv{} is approximate, the null distribution of its statistic is evaluated under independence, and we found that at the nominal 5\% $\alpha$-level, the type-I error for the ACAT was somewhat higher and could reach 7\% for some correlation settings. Nonetheless, the advantage of ACAT is that the approximation improves as the $\alpha$-level becomes smaller.

\subsection*{Simulations assuming that the correlation matrix is estimated using external data}
When only summary statistics are available, the correlation matrix $\matr{\Sigma}$ can be estimated from a reference panel of genotyped individuals. However, the type-I error of tests based on both $\text{TQ}$ and $\text{DOT}$ may potentially be affected due to substituting the sample estimate $\widehat{\matr{\Sigma}}$ by an estimate obtained from external data. To study the effect of this mis-specification on the type-I error, we conducted a separate set of simulations. In these experiments, we again utilized LD structures derived from the 1000 Genomes Project data. The type-I error rates, given in Table \ref{tab6}, show that both ACAT and $\text{TQ}$ have close to the nominal type-I error rates, but the error rate for the decorrelation method ($\text{DOT}$) can be inflated, unless the sample size of the reference panel is 50 to 100 times larger than the number of SNPs ($L$). For the statistic $\text{DOT}$, the type-I error rates appear to be more inflated at smaller $\alpha$-levels, such as $10^{-7}$ (according to preliminary simulations not shown here). Power values for $\text{TQ}$ are not shown, however they closely followed predicted theoretical power for the scenarios where the same data are used for both LD estimation and computation of association statistics. There was only 1 to 2\% drop in power when the size of the panel was only 2 to 5 times larger than $L$. 

\subsection*{Combining breast cancer association statistics within candidate genes}
We applied our decorrelation method to a family-based GWAS study of breast cancer.\cite{shi2017previous,o2016family} The data set was comprised of complete trios, i.e., families where genotypes of both parents and the affected offspring were available. With complete trios, previously reported statistics\cite{liu2010} become equivalent to statistics from the transmission-disequilibrium test and correlation among them is expected to follow the LD among SNPs.\cite{liu2010} We selected four candidate genes (\textit{TOX3, ESR1, FGFR2} and \textit{RAD51B}), for which Min et al.\cite{shi2017previous,o2016family} replicated several previously reported risk SNPs in relation to breast cancer.

\begin{figure}[th!]
\centering
 \includegraphics[width=0.4\linewidth]{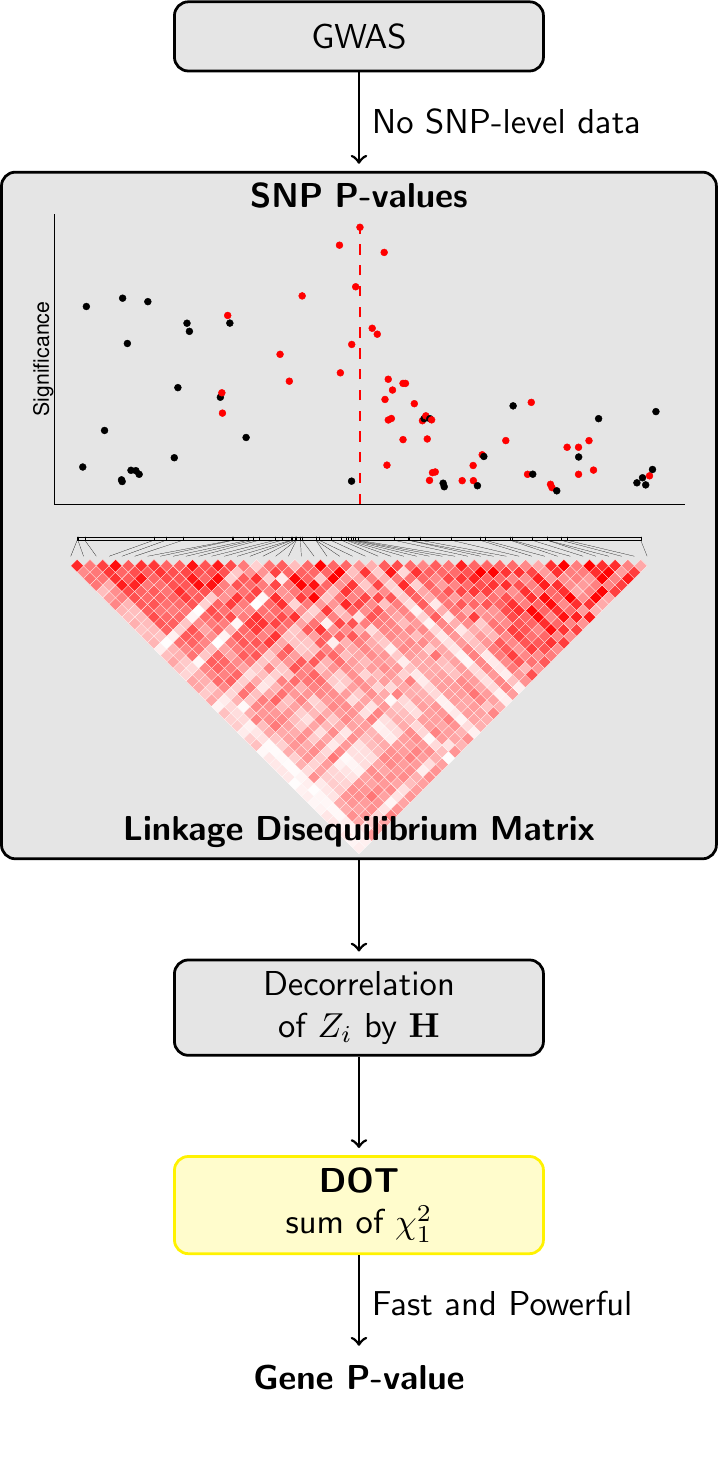} 
 \caption{\textbf{Overview of DOT method in application to breast cancer data}. We compute gene-level score by first decorrelating SNP P-values using the invariant to order matrix $\mathbf{H}$ and then calculating sum of independent chi-squared statistics. We utilize our DOT method to obtain a gene-level \Pv{}. In the breast cancer data application, we chose an anchor SNP -- a SNP that has previously been reported as risk variant (highlighted by a vertical dashed line), -- and then combine SNPs in an LD block with the anchor SNP by the DOT. SNP-level \Pvs{} highlighted in red are those in moderate to high LD with the anchor SNP.}
\label{fig:dot}
\end{figure}

For the joint association, we restricted our analysis to blocks of SNPs surrounding breast cancer risk variants that were previously reported in the literature. Specifically, we selected \textit{TOX3} rs4784220, \cite{ahsan2014genome} \textit{ESR1} rs3020314,\cite{lipphardt2013esr1,dunning2009association} \textit{FGFR2} rs2981579,\cite{ahsan2014genome} and \textit{RAD51B} rs999737,\cite{thomas2009multistage,michailidou2013large,pelttari2016rad51b}
and then included blocks of SNPs around these `anchor' risk variants with the LD correlation of at least 0.25. These blocks included 13 SNPs around rs4784220, 36 SNPs around rs3020314, 18 SNPs around rs2981579, and 30 SNPs around rs999737. As an illustration, Figure \ref{fig:dot} displays 81 SNP P-values that were available for \textit{ESR1} gene, the vertical dashed line highlights the position of `anchor' rs3020314, the red dots highlight 36 SNPs within LD-block of rs3020314, and the LD matrix displays sample correlation matrix among 36 SNPs. Once SNP blocks were identified for each gene, we applied four combination methods to assess their association with breast cancer.

Table \ref{tab7} present the joint association analysis results. The first row of Table \ref{tab7} shows P-values for the association between the LD block of 13 SNPs in \textit{TOX3} region and breast cancer. All methods conclude a statistically significant link but our decorrelation method provides the most robust evidence with a substantially lower P-value. The third row of Table \ref{tab7} shows joint association P-values for the LD block of 18 SNPs in \textit{FGFR2}. Three out of four methods conclude an association at 5\% level, with $\text{DOT}$ approach, once again, providing the most significant result. We note that the last column of Table \ref{tab7} gives the Bonferroni-style adjustment that is expected to be more conservative relative to the combination tests. Thus, it is not surprising that out of the four methods considered, the Bonferroni method failed to conclude an association. Lastly, the second and the fourth rows of Table \ref{tab7} provide joint association P-values for LD block in \textit{ESR1} and \textit{RAD51B}, respectively. For both \textit{ESR1} and \textit{RAD51B} our decorrelation approach was the only one that concluded a statistically significant association between SNP-set in those genes with breast cancer.

Table \ref{tab8} details a list of top SNPs that are associated with breast cancer within the selected candidate genes. The top ranked SNPs were identified by considering the top three components in the linear combination $\text{DOT} = \sum_{i=1}^L X_i^2$, where $X_i$'s are the decorrelated summary statistics. Once the highest three values of $X_i^2$ were identified for each gene, we considered individual components of $X_i = \sum_{j=1}^Lh_jZ_j$ that are formed as a linear combination of the original statistics weighted by the elements of matrix $\mathbf{H}$. The top individual components $h_jZ_j$ (with the same sign as $X_i$) were corresponding to individual SNPs presented in Table \ref{tab8}.

For the LD block in \textit{TOX3} gene, the top three individual $X_i$'s in $\text{DOT}$ statistic were all formed by having a very large weight assigned to a single SNP, i.e., the largest value, $X_{(1)}^2$, was formed by assigning a large weight to rs4784220 statistic; the second largest value, $X_{(2)}^2$,  was formed by assigning a large weight to rs8046979 statistic; and the third largest value, $X_{(3)}^2$, was formed by assigning a large weight to rs43143 statistic. The first few rows of Table \ref{tab8} detail these results and identify rs43143 as a new possible association with breast cancer.

For the LD block in \textit{ESR1} gene, the top $X_i$'s were quite different. Specifically, the largest value, $X_{(1)}$, was formed as a linear combination of 6 SNPs that all got assigned large weights. These 6 SNPs were rs2982689/rs3020424/rs985695/rs2347867/rs3003921/rs985191. The second highest linear combination, $X_{(2)}$, was formed by assigning high weights to 5 out of 6 SNPs listed above: rs2982689/rs3020424/rs985695/rs2347867/rs3003921. We note that the signs of $X_{(1)}$ and $X_{(2)}$ were in different directions and that is why it was possible for the same set of SNPs to be prioritized. Finally, the third largest value, $X_{(3)}$, also prioritized the same set of SNPs, with the exception of the single new addition of rs926777. Table \ref{tab8} provides a detailed discussion of these SNPs and identifies rs3003921/rs985695/rs2982689/rs3020424 and rs926777 as new possible associations with breast cancer.

Finally, for the LD blocks in \textit{FGFR2} and \textit{RAD51B} we repeated the procedure detailed above and also identified top-ranking SNPs. Table \ref{tab8} reviews these results and points \textit{FGFR2} rs2981427 and \textit{RAD51B} rs7359088 as two more additional newly found associations.

\section*{Discussion}
In this research, we have proposed a new powerful decorrelation-based approach ($\text{DOT}$) for combining SNP-level summary statistics (or, equivalently, \Pvs{}) and derived its theoretical power properties. To the best our knowledge, we were the first to derive theoretical properties of the traditional approach, $\text{TQ}$ (e.g., as implemented in VEGAS). Through extensive simulation studies, we have demonstrated that our decorrelation approach is a very powerful addition to the tools available for studying genetic susceptibility to disease.

Our analysis of breast cancer data illustrates unique properties of $\text{DOT}$. Our results revealed novel potential associations within four candidate genes that would have not been found by previously proposed methods. These novel SNPs were identified by examining the top three linear-combination contributors to the overall value of the DOT-statistic. We note that the top contributions may give large weights to genetic variants that are truly associated with the outcome or to SNPs in a high positive LD with true causal variants. Caution is needed when interpreting such results because our method cannot distinguish between causal and proxy associations. Further studies would be needed to confirm these findings.

The most important feature of the proposed method is that it provides substantial power boost across diverse settings, where power gain is amplified by heterogeneity of effect sizes and by increased diversity between pairwise LD values. Genetic architecture of complex traits is far from being homogeneous, making our method applicable in various settings.  We have developed new theory to explain unexpected and remarkable boost in power. This theory allows one to predict behavior of the tests in simulations with high accuracy and to explain unexpected scenarios, where the decorrelation method may give dramatically higher power compared to the traditional approach.  Yet, there are important precautions to the decorrelation approach. When reference panel data are used to provide the LD information and, more generally, correlation estimates for all predictors, including SNPs and covariates, $\hat{\matr{\Sigma}}$, sample size of the external data should be several times larger than the number of predictors. Ideally, the same data set should be used to obtain association statistics, as well as $\hat{\matr{\Sigma}}$. Nevertheless, association statistics and $\hat{\matr{\Sigma}}$ are compact summaries of data and are much more easily transferred between separate research groups than raw data, due to privacy considerations and potentially large size of the raw data sets. Also, caution is needed if missing data are present in the original data set because the estimate ($\hat{\matr{\Sigma}}$) may no longer reflect the sample correlation between predictors. Imputation of missing values is a suitable solution, if missing values are independent of the outcome. With the usage of reference panel data, the type-I error inflation for the statistic $\text{DOT}$ can be affected by many factors, and this statistic is expected to be sensitive not only to the size of a reference panel, but to population variations in LD, especially for highly correlated blocks of SNPs. Overall, it appears to be difficult to give specific recommendations, except that the reference panel size has to be at least 50 times larger than the number of SNPs to be combined. Therefore, we recommend to limit applications of the decorrelation method to situations, where the LD matrix is obtained from the same data set as the summary statistics. Note that all pairwise LD values can be obtained from sample haplotype frequencies of SNPs, thus the LD matrix can be reconstructed. Utility of this approach remains to be investigated, in particular, one concern is that the correlation between the SNP values reflect the composite disequilibrium values, \cite{zaykin2004bounds} while frequencies of sample haplotypes are often reported following likelihood maximization, e.g., by the EM algorithm.

In our simulations, the recently proposed method ACAT and the test based on the distribution of the sum of correlated association statistics (VEGAS, or $\text{TQ}$) had similar power. In many situations, power of these two tests was substantially lower than that of the $\text{DOT}$. The main advantage of ACAT is that it does not require any LD information. Our theory and simulations also revealed previously unknown robustness of the  $\text{TQ}$ method with respect to LD mis-specification: the method is valid and remains nearly as powerful when the sample LD matrix is substituted by a single value, summarizing the extent of all pairwise correlations. $\text{TQ}$ also remains valid when the LD summary is obtained from a representative reference panel of sample size as small as two to five times the number of SNPs to be combined. We stress again that compared to ACAT and $\text{TQ}$, our method's limitation is that in order to avoid possible bias, the LD information and the summary statistics should ideally come from the same data set and missing genotypes should be imputed prior to its application. In general, one should avoid utilization of external data as a source of LD information, as well as high rates of unimputed missing genotypes. Although not pursued here, a possible way to improve robustness of the $\text{DOT}$ is to merge it with ACAT, that is, decorrelate the summary statistics first, convert the results to \Pvs{} and then combine them with ACAT.

\section*{Declaration of Interests}
The authors declare no competing interests.

\section*{Acknowledgments}
This research was supported in part by the Intramural Research Program of the NIH, National Institute of Environmental Health Sciences.

\section*{Web Resources}
The URL for software referenced in this article is available at: \mbox{}\\
\noindent
{\small\url{https://github.com/dmitri-zaykin/Total_Decor}}

\clearpage
\bibliography{Total.Decor.04.19}

\begin{thebibliography}{10}

\bibitem{DanyuLinMetaNoGain2009}
Lin, D. and Zeng, D.
\newblock (2010).
\newblock {Meta-analysis of genome-wide association studies: No efficiency gain
  in using individual participant data}.
\newblock {Genet Epidemiol} {\em 34}, 60--66.

\bibitem{lee2013general}
Lee, S., Teslovich, T.~M., Boehnke, M., and Lin, X.
\newblock (2013).
\newblock General framework for meta-analysis of rare variants in sequencing
  association studies.
\newblock {Am J Hum Genet} {\em 93}, 42--53.

\bibitem{zaykin2011optimally}
Zaykin, D.~V.
\newblock (2011).
\newblock {Optimally weighted Z-test is a powerful method for combining
  probabilities in meta-analysis}.
\newblock J Evol Biol {\em 24}, 1836--1841.

\bibitem{pasaniuc2017dissecting}
Pasaniuc, B. and Price, A.~L.
\newblock (2017).
\newblock Dissecting the genetics of complex traits using summary association
  statistics.
\newblock Nature Reviews Genetics {\em 18}, 117.

\bibitem{Gates_ajhg}
Li, M.-X., Gui, H.-S., Kwan, J. S.~H., and Sham, P.~C.
\newblock (2011).
\newblock {GATES: a rapid and powerful gene-based association test using
  extended Simes procedure.}
\newblock {Am J Hum Genet} {\em 88}, 283--93.

\bibitem{conneely20071158}
Conneely, K.~N. and Boehnke, M.
\newblock (2007).
\newblock {So many correlated tests, so little time! Rapid adjustment of
  {P}-values for multiple correlated tests}.
\newblock The American Journal of Human Genetics {\em 81}, 1158--1168.

\bibitem{Sun361436}
Sun, R., Hui, S., Bader, G., Lin, X., and Kraft, P.
\newblock (2018).
\newblock {Powerful gene set analysis in GWAS with the generalized Berk-Jones
  statistic}.
\newblock bioRxiv, doi: https://doi.org/10.1101/361436.

\bibitem{liu2010}
Liu, J.~Z., Mcrae, A.~F., Nyholt, D.~R., Medland, S.~E., Wray, N.~R., Brown,
  K.~M., Hayward, N.~K., Montgomery, G.~W., Visscher, P.~M., Martin, N.~G.,
  et~al.
\newblock (2010).
\newblock A versatile gene-based test for genome-wide association studies.
\newblock {Am J Hum Genet} {\em 87}, 139--145.

\bibitem{lamparter2016fast}
Lamparter, D., Marbach, D., Rueedi, R., Kutalik, Z., and Bergmann, S.
\newblock (2016).
\newblock Fast and rigorous computation of gene and pathway scores from
  snp-based summary statistics.
\newblock PLoS computational biology {\em 12}, e1004714.

\bibitem{Zaykin2002}
Zaykin, D.~V., Zhivotovsky, L.~A., Westfall, P.~H., and Weir, B.~S.
\newblock (2002).
\newblock Truncated product method for combining {P}-values.
\newblock {Genet Epidemiol} {\em 22}, 170--85.

\bibitem{dudbridge2003}
Dudbridge, F. and Koeleman, B.~P.
\newblock (2003).
\newblock Rank truncated product of {P}-values, with application to genomewide
  association scans.
\newblock {Genet Epidemiol} {\em 25}, 360--366.

\bibitem{zaykin2007combining}
Zaykin, D.~V., Zhivotovsky, L.~A., Czika, W., Shao, S., and Wolfinger, R.~D.
\newblock (2007).
\newblock Combining {P}-values in large-scale genomics experiments.
\newblock Pharmaceutical Statistics: The Journal of Applied Statistics in the
  Pharmaceutical Industry {\em 6}, 217--226.

\bibitem{biernacka2012use}
Biernacka, J.~M., Jenkins, G.~D., Wang, L., Moyer, A.~M., and Fridley, B.~L.
\newblock (2012).
\newblock Use of the gamma method for self-contained gene-set analysis of {SNP}
  data.
\newblock European Journal of Human Genetics {\em 20}, 565.

\bibitem{fridley2013soft}
Fridley, B.~L., Jenkins, G.~D., Grill, D.~E., Kennedy, R.~B., Poland, G.~A.,
  and Oberg, A.~L.
\newblock (2013).
\newblock Soft truncation thresholding for gene set analysis of {RNA}-seq data:
  application to a vaccine study.
\newblock Scientific reports {\em 3}, 2898.

\bibitem{taylor2005tail}
Taylor, J. and Tibshirani, R.
\newblock (2005).
\newblock A tail strength measure for assessing the overall univariate
  significance in a dataset.
\newblock Biostatistics {\em 7}, 167--181.

\bibitem{maccallum2002practice}
MacCallum, R.~C., Zhang, S., Preacher, K.~J., and Rucker, D.~D.
\newblock (2002).
\newblock On the practice of dichotomization of quantitative variables.
\newblock Psychological methods {\em 7}, 19.

\bibitem{ferrari2012simulating}
Ferrari, P.~A. and Barbiero, A.
\newblock (2012).
\newblock Simulating ordinal data.
\newblock Multivariate Behavioral Research {\em 47}, 566--589.

\bibitem{clarkehelmert}
Clarke, B.~R.
\newblock (2008).
\newblock Helmert matrices and orthogonal relationships. In: Linear Models: The
  theory and application of analysis of variance.
\newblock (Wiley-Blackwell).

\bibitem{lancaster1965}
Lancaster, H.
\newblock (1965).
\newblock {The Helmert matrices}.
\newblock The American Mathematical Monthly {\em 72}, 4--12.

\bibitem{10002012integrated}
Consortium, . G.~P. et~al.
\newblock (2012).
\newblock An integrated map of genetic variation from 1,092 human genomes.
\newblock Nature {\em 491}, 56.

\bibitem{liu2019acat}
Liu, Y., Chen, S., Li, Z., Morrison, A.~C., Boerwinkle, E., and Lin, X.
\newblock (2019).
\newblock {ACAT: A fast and powerful P-value combination method for
  rare-variant analysis in sequencing studies}.
\newblock The American Journal of Human Genetics {\em 104}, 410--421.

\bibitem{wu2011rare}
Wu, M.~C., Lee, S., Cai, T., Li, Y., Boehnke, M., and Lin, X.
\newblock (2011).
\newblock Rare-variant association testing for sequencing data with the
  sequence kernel association test.
\newblock {Am J Hum Genet} {\em 89}, 82--93.

\bibitem{li2008methods}
Li, B. and Leal, S.~M.
\newblock (2008).
\newblock Methods for detecting associations with rare variants for common
  diseases: application to analysis of sequence data.
\newblock {Am J Hum Genet} {\em 83}, 311--321.

\bibitem{madsen2009groupwise}
Madsen, B.~E. and Browning, S.~R.
\newblock (2009).
\newblock A groupwise association test for rare mutations using a weighted sum
  statistic.
\newblock PLOS Genetics {\em 5}, e1000384.

\bibitem{price2010pooled}
Price, A.~L., Kryukov, G.~V., de~Bakker, P.~I., Purcell, S.~M., Staples, J.,
  Wei, L.-J., and Sunyaev, S.~R.
\newblock (2010).
\newblock Pooled association tests for rare variants in exon-resequencing
  studies.
\newblock The American Journal of Human Genetics {\em 86}, 832--838.

\bibitem{shi2017previous}
Shi, M., O’Brien, K.~M., Sandler, D.~P., Taylor, J.~A., Zaykin, D.~V., and
  Weinberg, C.~R.
\newblock (2017).
\newblock {Previous GWAS hits in relation to young-onset breast cancer}.
\newblock Breast cancer research and treatment {\em 161}, 333--344.

\bibitem{o2016family}
O'Brien, K.~M., Shi, M., Sandler, D.~P., Taylor, J.~A., Zaykin, D.~V., Keller,
  J., Wise, A.~S., and Weinberg, C.~R.
\newblock (2016).
\newblock A family-based, genome-wide association study of young-onset breast
  cancer: inherited variants and maternally mediated effects.
\newblock European Journal of Human Genetics {\em 24}, 1316.

\bibitem{ahsan2014genome}
Ahsan, H., Halpern, J., Kibriya, M.~G., Pierce, B.~L., Tong, L., Gamazon, E.,
  McGuire, V., Felberg, A., Shi, J., Jasmine, F., et~al.
\newblock (2014).
\newblock {A genome-wide association study of early-onset breast cancer
  identifies PFKM as a novel breast cancer gene and supports a common genetic
  spectrum for breast cancer at any age}.
\newblock Cancer Epidemiology and Prevention Biomarkers {\em 23}, 658--669.

\bibitem{lipphardt2013esr1}
Lipphardt, M.~F., Deryal, M., Ong, M.~F., Schmidt, W., and Mahlknecht, U.
\newblock (2013).
\newblock {ESR1 single nucleotide polymorphisms predict breast cancer
  susceptibility in the central European Caucasian population}.
\newblock International journal of clinical and experimental medicine {\em 6},
  282.

\bibitem{dunning2009association}
Dunning, A.~M., Healey, C.~S., Baynes, C., Maia, A.-T., Scollen, S., Vega, A.,
  Rodr{\'\i}guez, R., Barbosa-Morais, N.~L., Ponder, B.~A., Low, Y.-L., et~al.
\newblock (2009).
\newblock {Association of ESR1 gene tagging SNPs with breast cancer risk}.
\newblock Human molecular genetics {\em 18}, 1131--1139.

\bibitem{thomas2009multistage}
Thomas, G., Jacobs, K.~B., Kraft, P., Yeager, M., Wacholder, S., Cox, D.~G.,
  Hankinson, S.~E., Hutchinson, A., Wang, Z., Yu, K., et~al.
\newblock (2009).
\newblock {A multistage genome-wide association study in breast cancer
  identifies two new risk alleles at 1p11. 2 and 14q24. 1 (RAD51L1)}.
\newblock Nature genetics {\em 41}, 579.

\bibitem{michailidou2013large}
Michailidou, K., Hall, P., Gonzalez-Neira, A., Ghoussaini, M., Dennis, J.,
  Milne, R.~L., Schmidt, M.~K., Chang-Claude, J., Bojesen, S.~E., Bolla, M.~K.,
  et~al.
\newblock (2013).
\newblock Large-scale genotyping identifies 41 new loci associated with breast
  cancer risk.
\newblock Nature genetics {\em 45}, 353.

\bibitem{pelttari2016rad51b}
Pelttari, L.~M., Khan, S., Vuorela, M., Kiiski, J.~I., Vilske, S., Nevanlinna,
  V., Ranta, S., Schleutker, J., Winqvist, R., Kallioniemi, A., et~al.
\newblock (2016).
\newblock {RAD51B in familial breast cancer}.
\newblock PloS one {\em 11}, e0153788.

\bibitem{zaykin2004bounds}
Zaykin, D.~V.
\newblock (2004).
\newblock Bounds and normalization of the composite linkage disequilibrium
  coefficient.
\newblock {Genet Epidemiol} {\em 27}, 252--257.

\bibitem{udler2010fine}
Udler, M.~S., Ahmed, S., Healey, C.~S., Meyer, K., Struewing, J., Maranian, M.,
  Kwon, E.~M., Zhang, J., Tyrer, J., Karlins, E., et~al.
\newblock (2010).
\newblock Fine scale mapping of the breast cancer 16q12 locus.
\newblock Human molecular genetics {\em 19}, 2507--2515.

\bibitem{linjawi2019relation}
Linjawi, S.~A., Hifni, S.~A., and ALKhayyat, S.~S.
\newblock (2019).
\newblock The relation between estrogen-positive receptor in breast cancer
  ({ER+}) and obesity in jeddah.
\newblock Journal of Biology and Today's World {\em 8}, 13--20.

\bibitem{sonestedt2009protective}
Sonestedt, E., Ivarsson, M.~I., Harlid, S., Ericson, U., Gullberg, B., Carlson,
  J., Olsson, H., Adlercreutz, H., and Wirfalt, E.
\newblock (2009).
\newblock The protective association of high plasma enterolactone with breast
  cancer is reasonably robust in women with polymorphisms in the estrogen
  receptor $\alpha$ and $\beta$ genes.
\newblock The Journal of nutrition {\em 139}, 993--1001.

\bibitem{yingchun2009relationship}
Yingchun, X., Zhang, F., Wang, H., Ma, Y., and Sun, L.
\newblock (2009).
\newblock Relationship between single nucleotide polymorphism of estrogen
  receptor gene and endocrine therapy efficacy in breast cancer.
\newblock Journal of Clinical Oncology {\em 27}, 1113--1113.

\bibitem{nyante2015genetic}
Nyante, S.~J., Gammon, M.~D., Kaufman, J.~S., Bensen, J.~T., Lin, D.~Y.,
  Barnholtz-Sloan, J.~S., Hu, Y., He, Q., Luo, J., and Millikan, R.~C.
\newblock (2015).
\newblock Genetic variation in estrogen and progesterone pathway genes and
  breast cancer risk: an exploration of tumor subtype-specific effects.
\newblock Cancer Causes \& Control {\em 26}, 121--131.

\bibitem{mahoney2013predicting}
Mahoney, D.~W., Kohli, M., Cerhan, J.~R., and Offer, S.~M. (2013).
\newblock Predicting responses to androgen deprivation therapy.
\newblock US Patent App. 13/821,807.

\bibitem{saadatian2014association}
Saadatian, Z., Gharesouran, J., Ghojazadeh, M., Ghohari-Lasaki, S.,
  Tarkesh-Esfahani, N., and Ardebili, S. M.~M.
\newblock (2014).
\newblock Association of rs1219648 in {FGFR2} and rs1042522 in {TP53} with
  premenopausal breast cancer in an iranian azeri population.
\newblock Asian Pacific Journal of Cancer Prevention {\em 15}, 7955--7958.

\bibitem{andersen2013breast}
Andersen, S.~W., Trentham-Dietz, A., Figueroa, J.~D., Titus, L.~J., Cai, Q.,
  Long, J., Hampton, J.~M., Egan, K.~M., and Newcomb, P.~A.
\newblock (2013).
\newblock Breast cancer susceptibility associated with rs1219648 (fibroblast
  growth factor receptor 2) and postmenopausal hormone therapy use in a
  population-based {United States} study.
\newblock Menopause (New York, NY) {\em 20}, 354--358.

\bibitem{zhang2017association}
Zhang, Y., Zeng, X., Liu, P., Hong, R., Lu, H., Ji, H., Lu, L., and Li, Y.
\newblock (2017).
\newblock {Association between FGFR2 (rs2981582, rs2420946 and rs2981578)
  polymorphism and breast cancer susceptibility: a meta-analysis}.
\newblock Oncotarget {\em 8}, 3454.

\bibitem{zhang2010current}
Zhang, J., Qiu, L.-X., Wang, Z.-H., Leaw, S.-J., Wang, B.-Y., Wang, J.-L., Cao,
  Z.-G., Gao, J.-L., and Hu, X.-C.
\newblock (2010).
\newblock Current evidence on the relationship between three polymorphisms in
  the {FGFR2} gene and breast cancer risk: a meta-analysis.
\newblock Breast cancer research and treatment {\em 124}, 419--424.

\bibitem{chen2011risk}
Chen, X.-H., Li, X.-Q., Chen, Y., and Feng, Y.-M.
\newblock (2011).
\newblock Risk of aggressive breast cancer in women of han nationality carrying
  {TGFB1} rs1982073 c allele and {FGFR2} rs1219648 g allele in north china.
\newblock Breast cancer research and treatment {\em 125}, 575--582.

\bibitem{lei2017fibroblast}
Lei, H. and Deng, C.-X.
\newblock (2017).
\newblock Fibroblast growth factor receptor 2 signaling in breast cancer.
\newblock International journal of biological sciences {\em 13}, 1163.

\bibitem{murillo2013association}
Murillo-Zamora, E., Moreno-Mac{\'\i}as, H., Ziv, E., Romieu, I., Lazcano-Ponce,
  E., {\'A}ngeles-Llerenas, A., P{\'e}rez-Rodr{\'\i}guez, E., Vidal-Mill{\'a}n,
  S., Fejerman, L., and Torres-Mej{\'\i}a, G.
\newblock (2013).
\newblock Association between rs2981582 polymorphism in the {FGFR2} gene and
  the risk of breast cancer in mexican women.
\newblock Archives of medical research {\em 44}, 459--466.

\bibitem{butt2012genetic}
Butt, S., Harlid, S., Borgquist, S., Ivarsson, M., Landberg, G., Dillner, J.,
  Carlson, J., and Manjer, J.
\newblock (2012).
\newblock Genetic predisposition, parity, age at first childbirth and risk for
  breast cancer.
\newblock BMC research notes {\em 5}, 414.

\bibitem{shan2012genome}
Shan, J., Mahfoudh, W., Dsouza, S.~P., Hassen, E., Bouaouina, N., Abdelhak, S.,
  Benhadjayed, A., Memmi, H., Mathew, R.~A., Aigha, I.~I., et~al.
\newblock (2012).
\newblock {Genome-Wide Association Studies (GWAS) breast cancer susceptibility
  loci in Arabs: susceptibility and prognostic implications in Tunisians}.
\newblock Breast cancer research and treatment {\em 135}, 715--724.

\bibitem{xu2011relation}
Xu, W.-H., Shu, X.-O., Long, J., Lu, W., Cai, Q., Zheng, Y., Xiang, Y.-B., Dai,
  Q., Zhao, G.-m., Gu, K., et~al.
\newblock (2011).
\newblock {Relation of FGFR2 genetic polymorphisms to the association between
  oral contraceptive use and the risk of breast cancer in Chinese women}.
\newblock American journal of epidemiology {\em 173}, 923--931.

\bibitem{dong2014analyzing}
Dong, H., Gao, Z., Li, C., Wang, J., Jin, M., Rong, H., Niu, Y., and Liu, J.
\newblock (2014).
\newblock Analyzing 395,793 samples shows significant association between
  rs999737 polymorphism and breast cancer.
\newblock Tumor Biology {\em 35}, 6083--6087.

\bibitem{turnbull2010genome}
Turnbull, C., Ahmed, S., Morrison, J., Pernet, D., Renwick, A., Maranian, M.,
  Seal, S., Ghoussaini, M., Hines, S., Healey, C.~S., et~al.
\newblock (2010).
\newblock Genome-wide association study identifies five new breast cancer
  susceptibility loci.
\newblock Nature genetics {\em 42}, 504.

\bibitem{lee2012fine}
Lee, P., Fu, Y.-P., Figueroa, J.~D., Prokunina-Olsson, L., Gonzalez-Bosquet,
  J., Kraft, P., Wang, Z., Jacobs, K.~B., Yeager, M., Horner, M.-J., et~al.
\newblock (2012).
\newblock Fine mapping of 14q24. 1 breast cancer susceptibility locus.
\newblock Human genetics {\em 131}, 479--490.

\bibitem{stacey2015genetic}
Stacey, S. and Sulem, P. (2015).
\newblock Genetic variants for breast cancer risk assessment.
\newblock US Patent 8,951,735.

\bibitem{ma2012genetic}
Ma, H., Li, H., Jin, G., Dai, J., Dong, J., Qin, Z., Chen, J., Wang, S., Wang,
  X., Hu, Z., et~al.
\newblock (2012).
\newblock {Genetic variants at 14q24. 1 and breast cancer susceptibility: a
  fine-mapping study in Chinese women}.
\newblock DNA and cell biology {\em 31}, 1114--1120.

\end{thebibliography}

\clearpage
\section*{Figure Titles and Legends}
\clearpage
\section*{Tables}
\begin{table}[th]
  \centering
\begin{tabular}{llllllll}
  \hline \\
  Number of SNPs &  Empiric. & Theor.  & Approx.  & Empiric.   & Theor.  & ACAT & $\bar{\gamma}$  \\
  $L$  & $\text{TQ}$ & $\text{TQ}$ & $\text{TQ}$ & $\text{DOT}$  & $\text{DOT}$  & &  \\ \hline
 500  &  0.802   &  0.802    &  0.802       &    0.090    &     0.090  & 0.832    &     0.02    \\
 300  &  0.801   &  0.801    &  0.801       &    0.101    &     0.100  & 0.830    &     0.03    \\
 200  &  0.801   &  0.801    &  0.801       &    0.112    &     0.112  & 0.829    &     0.04    \\
 100  &  0.799   &  0.800    &  0.800       &    0.144    &     0.145  & 0.826    &     0.08    \\
 50   &  0.798   &  0.799    &  0.799       &    0.196    &     0.197  & 0.821    &     0.16     \\
 30   &  0.795   &  0.796    &  0.796       &    0.253    &     0.252  & 0.814    &     0.26     \\
 20   &  0.794   &  0.793    &  0.794       &    0.307    &     0.306  & 0.809    &     0.39    \\
 \hline
\end{tabular}
\caption{Power comparison of $\text{TQ}$, $\text{DOT}$, and ACAT, assuming very similar effect sizes in magnitude and equicorrelation LD structure with $\rho$=0.7.}
\label{tab1}
\end{table}

\begin{table}[th]
  \centering
\begin{tabular}{llllllll}
  \hline \\
  Number of SNPs &  Empiric. & Theor.  & Approx.  & Empiric.   & Theor. & ACAT  & $\bar{\gamma}$\\
  $L$  & $\text{TQ}$ & $\text{TQ}$ & $\text{TQ}$ & $\text{DOT}$ & $\text{DOT}$  & &  \\ \hline
 500   &  0.729   &    0.730    &   0.726    &     0.973     &       0.973  & 0.793   &        0.251      \\
 300   &  0.731   &    0.730    &   0.726    &     0.883     &       0.883  & 0.791   &        0.256      \\
 200   &  0.731   &    0.730    &   0.726    &     0.810     &       0.811  & 0.789   &        0.281      \\
 100   &  0.730   &    0.731    &   0.726    &     0.599     &       0.599  & 0.786   &        0.295      \\
 50    &  0.732   &    0.733    &   0.728    &     0.577     &       0.576  & 0.782   &        0.418      \\
 30    &  0.736   &    0.735    &   0.729    &     0.504     &       0.502  & 0.778   &        0.488      \\
 20    &  0.737   &    0.737    &   0.731    &     0.541     &       0.540  & 0.776   &        0.661      \\
 \hline
\end{tabular}
\caption{Power comparison of $\text{TQ}$, $\text{DOT}$, and ACAT, assuming very similar effect sizes but heterogeneous LD structure.}
\label{tab2}
\end{table}

\begin{table}[th]
  \centering
\begin{tabular}{llllllll}
  \hline \\
  Number of SNPs &  Empiric. & Theor.  & P-approx.  & Empiric.   & Theor.  & ACAT & $\bar{\gamma}$  \\
  $L$  & $\text{TQ}$ & $\text{TQ}$ & $\text{TQ}$ & $\text{DOT}$ & $\text{DOT}$  & &  \\ \hline
 500   &  0.525    &   0.525   &     0.526    &      1.000    &        1.000  & 0.626   &      0.479     \\
 300   &  0.526    &   0.525   &     0.526    &      1.000    &        0.999  & 0.624   &      0.486     \\
 200   &  0.526    &   0.525   &     0.524    &      0.993    &        0.993  & 0.622   &      0.494     \\
 100   &  0.525    &   0.524   &     0.524    &      0.919    &        0.920  & 0.616   &      0.518     \\
 50    &  0.522    &   0.523   &     0.522    &      0.762    &        0.762  & 0.607   &      0.566     \\
 30    &  0.521    &   0.521   &     0.521    &      0.648    &        0.648  & 0.599   &      0.630    \\
 20    &  0.519    &   0.519   &     0.520    &      0.578    &        0.579  & 0.592   &      0.709    \\
 \hline
\end{tabular}
\caption{Power comparison of $\text{TQ}$, $\text{DOT}$, and ACAT, assuming heterogeneity in effect sizes but equicorrelated LD.}
\label{tab3}
\end{table}

\begin{table}[th]
  \centering
\begin{tabular}{llllllll}
  \hline \\
  Number of SNPs &  Empiric. & Theor.  & P-approx.  & Empiric.   & Theor. & ACAT & $\bar{\gamma}$  \\
  $L$  & $\text{TQ}$ & $\text{TQ}$ & $\text{TQ}$ & $\text{DOT}$ & $\text{DOT}$  & &  \\ \hline
 500 & 0.0500 &  0.0503 &  0.0508    & 0.9226    &    0.9222   &  0.0564 & 0.2118     \\
 300 & 0.0506 &  0.0503 &  0.0509    & 0.7688    &    0.7689   &  0.0570 & 0.2107     \\
 200 & 0.0504 &  0.0503 &  0.0508    & 0.5970    &    0.5967   &  0.0570 & 0.2025     \\
 100 & 0.0504 &  0.0503 &  0.0509    & 0.3040    &    0.3038   &  0.0568 & 0.1655     \\
 50  & 0.0502 &  0.0503 &  0.0508    & 0.3074    &    0.3070   &  0.0555 & 0.2397     \\
 30  & 0.0505 &  0.0503 &  0.0507    & 0.1485    &    0.1487   &  0.0562 & 0.1527     \\
 20  & 0.0501 &  0.0503 &  0.0508    & 0.1191    &    0.1189   &  0.0557 & 0.1399     \\
 \hline
\end{tabular}
\caption{Power comparison of $\text{TQ}$, $\text{DOT}$, and ACAT with effect sizes randomly sampled from -0.15 to 0.15 and heterogeneous LD.}
\label{tab4}
\end{table}

\clearpage
\begin{table}[th]
  \centering
\begin{tabular}{cccccccccc}
  \hline
 & Theor.  & Approx. & Regr. & MVN  & Theor. & Regr. & MVN & \\
 & $\text{TQ}$  & $\text{TQ}$ & $\text{TQ}$ & $\text{TQ}$ & $\text{DOT}$ & $\text{DOT}$ & $\text{DOT}$ & ACAT \\
  \hline 

Setting 5 &  &  &  & &   &  &  &        \\
$\alpha=10^{-3}$ & 0.34 &  0.34 & 0.34 & 0.34 & 0.60 &  0.60 &  0.60 & 0.40 \\
Setting 6 &  &  &  & &   &  &  &        \\
$\alpha=10^{-4}$ & 0.42 &  0.42 & 0.42 & 0.43 & 0.77 &  0.77 &  0.77 & 0.43 \\
Setting 7 &  &  &  & &   &  &  &        \\
$\alpha=10^{-7}$ & 0.24 &  0.24 & 0.24 & 0.24 & 0.76 &  0.76 &  0.76 & 0.18 \\
  \hline
\end{tabular}
\caption{Power comparison of $\text{TQ}$, $\text{DOT}$, and ACAT using realistic LD patterns from 1000 Genomes project. \\
}
\label{tab5}
\end{table}

\begin{table}[th]
  \centering
\begin{tabular}{cccc}
  \hline
 Sample size & $\text{TQ}$ & $\text{DOT}$ & ACAT \\
  \hline 

$N = 5L$    & $9\times 10^{-5}$    & $5\times 10^{-4}$  &  $1\times 10^{-4}$       \\
$N = 10L$    & $9\times 10^{-5}$    & $4\times 10^{-4}$  &  $1\times 10^{-4}$       \\
$N = 50L$   & $1\times 10^{-4}$    & $1\times 10^{-4}$  &  $1\times 10^{-4}$       \\
$N = 100L$    & $1\times 10^{-4}$    & $1\times 10^{-4}$  &  $1\times 10^{-4}$       \\
  \hline
\end{tabular}
\caption{Type-I error rates ($\alpha = 10^{-4}$) using a reference panel to estimate LD. Population LD patterns are modeled using 1000 Genomes project data.}
\label{tab6}
\end{table}

\begin{table}[th]
  \centering
\begin{tabular}{lcccc}
  \hline
  Gene & $\text{TQ}$ & $\text{DOT}$ & ACAT & $\min(P)\times L$ \\
  \hline
 \textit{TOX3}/rs4784220\cite{ahsan2014genome} ($L=13$) & 0.0005 & 0.0004   & 0.001 & 0.001 \\
 
  \textit{ESR1}/rs3020314\cite{lipphardt2013esr1,dunning2009association} ($L=36$) & 0.20   & 0.0001 & 0.19 & 0.96 \\ 
 
  \textit{FGFR2}/rs2981579\cite{ahsan2014genome} ($L=18$) & 0.01 & 0.003 & 0.01 & 0.07 \\
 
  \textit{RAD51B}/rs999737\cite{thomas2009multistage,michailidou2013large,pelttari2016rad51b} ($L=30$) & 0.56 & 0.009 & 0.76 & 1 \\
  \hline
\end{tabular}
\caption{Breast cancer candidate gene association \Pvs{}.}
\label{tab7}
\end{table}

\clearpage
\begin{table}[th]
  \centering
\begin{tabular}{l p{2cm} l p{9cm}}
  \hline
  Gene & Number of SNPs in analysis ($L$) & rs number & Reference \\
  \hline
  TOX3 & 13 & rs4784220 & This SNP was previously reported in the literature to be associated with breast cancer.\cite{ahsan2014genome, udler2010fine}  \\
                  &   & rs8046979 & This SNP was also linked to breast cancer.\cite{ahsan2014genome}\\
                  &   & \textbf{rs43143} & A new association with susceptibility to breast cancer.\\ \hline 
  ESR1 & 36 & rs2347867 & This SNP was previously reported to be involved in breast cancer risk. \cite{linjawi2019relation, sonestedt2009protective} \\
       & & rs985191 & This SNP was previously reported to be associated with endocrine therapy efficacy in breast cancer, \cite{yingchun2009relationship} as well as with the overall breast cancer risk.\cite{nyante2015genetic} \\
       & & \textbf{rs3003921} & A new association with susceptibility to breast cancer. This SNP was previously linked to the effectiveness of androgen deprivation therapy among prostate cancer patients. \cite{mahoney2013predicting} \\
       & & \textbf{rs985695} & A new association with susceptibility to breast cancer.\\
       & & \textbf{rs2982689} & A new association with susceptibility to breast cancer. \\
       & & \textbf{rs3020424} & A new association with susceptibility to breast cancer. \\
       & & \textbf{rs926777} & A new association with susceptibility to breast cancer.\\\hline
  FGFR2 & 18 & rs1219648 & This SNP was previously reported to be associated with premenopausal breast cancer\cite{saadatian2014association} and the overall breast cancer risk.\cite{andersen2013breast, zhang2017association, zhang2010current, chen2011risk} \\
       & & rs2860197 & This SNP was previously suggested to have an association with breast cancer.\cite{lei2017fibroblast} \\
       & & rs2981582 & This SNP was previously reported in the literature to be associated with breast cancer.\cite{murillo2013association, butt2012genetic, shan2012genome, zhang2017association}\\
       & & rs3135730 & This SNP was previously suggested to have an interaction between oral contraceptive use and breast cancer.\cite{xu2011relation}\\
       & & \textbf{rs2981427} & A new association with susceptibility to breast cancer.\\\hline
  RAD51B & 30 & rs999737 & This SNP was previously reported in the literature to be associated with breast cancer.\cite{dong2014analyzing, thomas2009multistage, turnbull2010genome} \\
       & & rs8016149 &  This SNP was previously suggested to have an association with breast cancer.\cite{lee2012fine} \\
       & & rs999737 & This SNP was previously reported in the literature to be associated with breast cancer.7\cite{thomas2009multistage,michailidou2013large,pelttari2016rad51b} \\
       & & rs1023529 & This SNP has been patented as one of susceptibility variants of breast cancer.\cite{stacey2015genetic}\\
       & & rs2189517 & This SNP was showed to be associated with breast cancer in Chinese population.\cite{ma2012genetic}\\
   & & \textbf{rs7359088} &  A new association with susceptibility to breast cancer.\\\hline
\end{tabular}
\caption{Breast cancer SNPs identified by $\text{DOT}$ in the analysis of GWAS data.}
\label{tab8}
\end{table}


\end{document}